\documentclass[journal]{IEEEtran}

\usepackage{amsmath}
\usepackage{cite}
\usepackage{mdwtab}
\usepackage{subfigure}
\usepackage{placeins}
\usepackage{psfrag}
\usepackage{graphicx}
\usepackage{epsfig}
\usepackage{epstopdf}
\usepackage[latin1]{inputenc}
\usepackage{amssymb}
\usepackage{amsthm}
\usepackage{comment}

\theoremstyle{plain}
\newtheorem{theorem}{Theorem}

\theoremstyle{remark}
\newtheorem{lemma}{Lemma}

\IEEEoverridecommandlockouts 

\begin{document}

\title{Scaling Laws of the Throughput Capacity and Latency in Information-Centric Networks}

\author{\IEEEauthorblockN{Bita Azimdoost$^\dagger$\thanks{Bita Azimdoost was with Huawei Innovation Center, Santa Clara, CA 95050, USA, as an intern while working on this paper.}, Cedric Westphal$^\ddagger$ $^*$, and Hamid R. Sadjadpour$^\dagger$}\\
\IEEEauthorblockA{$^\dagger$Department of Electrical Engineering and $^\ddagger$Computer Engineering\\
University of California Santa Cruz, Santa Cruz, CA 95064, USA\\
\{bazimdoost,cedric,hamid\}@soe.ucsc.edu\\
$^*$ Huawei Innovation Center, Santa Clara, CA 95050, USA\\
cedric.westphal@huawei.com}}

\maketitle

\begin{abstract}
Wireless information-centric networks consider storage as one of the network primitives, and propose to cache data within the network in order to improve latency and reduce bandwidth consumption. We study the throughput capacity and delay in an information-centric network when the data cached in each node has a limited lifetime. The results show that with some fixed request and cache expiration rates, the order of the data access time does not change with network growth, and the maximum throughput order is inversely proportional to the square root and logarithm of the network size $n$ in cases of grid and random networks, respectively. Comparing these values with the corresponding throughput and latency with no cache capability (throughput inversely proportional to the network size, and latency of order $\sqrt{n}$ and $\sqrt{\frac{n}{\log n}}$ in grid and random networks, respectively), we can actually quantify the asymptotic advantage of caching. Moreover, we compare these scaling laws for different content discovery mechanisms and illustrate that not much gain is lost when a simple path search is used. 
\end{abstract}

\IEEEpeerreviewmaketitle

\section{Introduction}

In today's networking situations, users are mostly interested in accessing content regardless of which host is providing this content. They are looking for a fast and secure access to data in a whole range of situations: wired or wireless; heterogeneous technologies; in a fixed location or when moving. The dynamic characteristics of the network users makes the host-centric networking paradigm inefficient. Information-centric networking (ICN) is a new networking architecture where content is accessed based upon its name, and independently of the location of the hosts \cite{Zhang2010Named,Pursuit,Ahlgren2012Survey,Jacobson2009Networking}. In most ICN architectures, data is allowed to be stored in the nodes and routers within the network in addition to the content publisher's servers. This reduces the burden on the servers and on the network operator, and shortens the access time to the desired content.

Combining content routing with in-network-storage for the information is intuitively attractive, but there has been few works considering the impact of such architecture on the capacity of the network in a formal or analytical manner. In this work we study a wireless information-centric network where nodes can both route and cache content\footnote{A preliminary version of this paper has appeared at ITC25 \cite{Azimdoost2013Throughput}}. We also assume that a node will keep a copy of the content only for a finite period of time, that is until it runs out of memory space in its cache and has to rotate content, or until it ceases to serve a specific content.

The nodes issue some queries for content that is not locally available. We suppose that there exists a server which permanently keeps all the contents. This means that the content is always provided at least by its publisher, in addition to the potential copies distributed throughout the network. Therefore, at least one replica of each content always exists in the network and if a node requests a piece of information, this data will be provided either by its original server or by a cache containing the desired data. When the customer receives the content, it will store the content and share it with the other nodes if needed.

The present paper thus investigates the access time and throughput capacity in such content-centric networks and addresses the following questions:

\begin{enumerate}
	\item Looking at the throughput capacity and latency, can we quantify the performance improvement brought about by a  content-centric network architecture over networks with no content sharing capability?
	\item How does the content discovery mechanism affect the performance? More specifically, does selecting the nearest copy of the content improve the scaling of the capacity and access time compared to selecting the nearest copy in the direction of original server?  
	\item How does the caching policy, and in particular, the length of time each piece of content spends in the cache's memory, affect the performance?
\end{enumerate}

We state our results in three Theorems; Theorem \ref{thm:01} formulates the throughput capacity in a grid network which uses the shortest path to the server content discovery mechanism considering different content availability in different caches, and Theorem \ref{thm:02} and \ref{thm:03} will answer the above questions studying two different network models (grid and random network) and two content discovery scenarios (shortest path to the server and shortest path to the closest copy of the content) when the information exists in all caches with the same probability. 
These Theorems demonstrate that adding the content sharing capability to the nodes can significantly increase the capacity.

The rest of the paper is organized as follows. After a brief review of the related work in Section \ref{sec:related}, the network models, the content discovery algorithms used in the current work, and the content distribution in steady-state are introduced in Section \ref{sec:netmodel}. The main Theorems are stated and proved in Section \ref{sec:theorems}. We will discuss the results and study some simple examples in Section \ref{sec:discussion}. Finally the paper is concluded and some possible directions for the future work will be introduced in section \ref{sec:conclusion}.

\section{Related Work}
\label{sec:related}

Information Centric Networks have recently received considerable attention. While our work presents an analytical abstraction, it is based upon the principles described in some ICN architectures, such as CCN~\cite{Jacobson2009Networking}, NetInf~\cite{Ahlgren2008Design}, PURSUIT~\cite{Pursuit}, or DONA~\cite{Koponen2007Dataoriented}, where nodes can cache content, and requests for content can be routed to the nearest copy. Papers surveying the landscape of ICN~\cite{Ahlgren2012Survey}\cite{Ghodsi2011InformationCentric} show the dearth of theoretical results underlying these architectures. 

Caching, one of the main concepts in ICN networks, has been studied in prior works~\cite{Ahlgren2012Survey}. \cite{Olmos2014Catalog} computes the performance of a LRU cache taking into account the dynamical nature of the content catalog. Some performance metrics like miss ratio in the cache, or the average number of hops each request travels to locate the content have been studied in \cite{InfoCom01:Che,InfoCom10:Rosensweig}, and the benefit of cooperative caching has been investigated in \cite{Wolman1999Scale}. 

Optimal cache locations \cite{Rosensweig2009Breadcrumbs} and cache replacement techniques \cite{IEEEMob05:Yin} are two other aspects most commonly investigated. And an analytical framework for investigating properties of these networks like fairness of cache usage is proposed in \cite{Tortelli2011Fairness}. \cite{Westphal2005Maximizing} considered information being cached for a limited amount of time at each node, as we do here, but focused on flooding mechanism to locate the content, not on the capacity of the network. \cite{Dehghan2014Complexity} investigates the routing in such networks in order to minimize the average access delay.

However, to the best of our knowledge, there are just a few works focusing on the achievable data rates in such networks. Calculating the asymptotic throughput capacity of wireless networks with no cache has been solved in~\cite{Gupta2000Capacity} and many subsequent works~\cite{Li2001Capacity}\cite{Niesen2009Capacity}. Some work has studied the capacity of wireless networks with caching \cite{Grossglauser2002Mobility}\cite{Herdtner2005Throughput}\cite{AlfanoContentCentric} . There, caching is used to buffer data at a relay node which will physically move to deliver the content to its destination, whereas we follow the ICN assumption that caching is triggered by the node requesting the content. 
\cite{ICCW09:Liu} uses a network simulation model and evaluates the performance (file transfer delay) in a cache-and-forward system with no request for the data. \cite{Carofiglio2011Modeling} proposes an analytical model for single cache miss probability and stationary throughput in cascade and binary tree topologies. \cite{IEEEIT11:Niesen} considers a general problem of delivering content cached in a wireless network and provides some bounds on the caching capacity region from an information-theoretic point of view. Some scaling regimes for the required link capacity is computed in \cite{Gitzenis2013Asymptotic} for a static cache placement in a multihop wireless network.

\section{Preliminaries}
\label{sec:netmodel}

\subsection{Network Model}
Two network models are studied in this work.
\subsubsection{Grid Network}
\begin{figure}[http]
    \center
      \includegraphics[scale=0.4,angle=0]{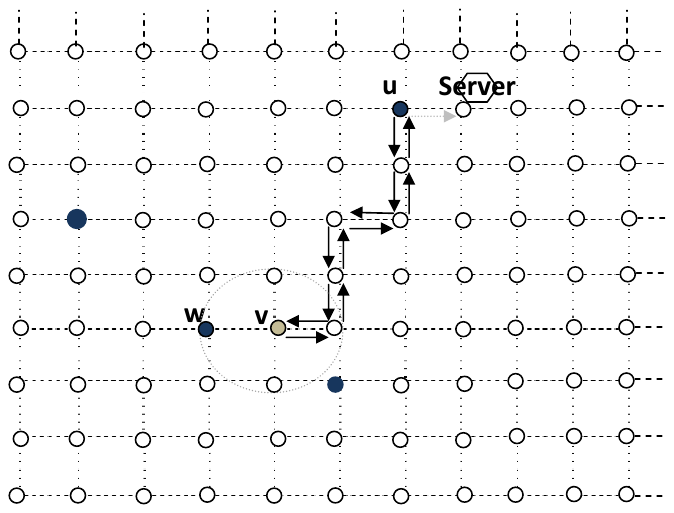}\\
      \caption{The transmission range of node $v$ contains four surrounding nodes. The black vertices contain the content in their  local caches. The arrow lines demonstrate a possible discovery and receive path in scenario $\romannumeral 1$, where node $v$ downloads the required information from $u$. In scenario $\romannumeral 2$, $v$ will download the data from $w$ instead.}
    \label{fig:gridnet}
\end{figure}

Assume that the network consists of $n$ nodes $V=\{v_1,v_2,...,v_n\}$ each with a local cache of size $L$ located on a grid (Figure \ref{fig:gridnet}). The distance between two adjacent nodes equals to the transmission range of each node, so the packets sent from a node are only received by four adjacent nodes. There are $m$ different contents, $F=\{f_1,f_2,...,f_m\}$ with sizes $B_i,\ i=1,...,m$, for which each node $v_j$ may issue a query. Based on the content discovery algorithms which will be explained later in this section, the query will be transmitted in the network to discover a node containing the desired content locally. $v_j$ then downloads $b$ bits of data with rate $\gamma$ in a hop-by-hop manner through the path $P_{xj}$ from either a node ($v_i, x=i$) containing it locally ($f\in v_i$) or the server ($x=s$). When the download is completed, the data is cached and shared with other nodes either by all the nodes on the delivery path, or only by the end node. In the paper we consider both options. 

$P_{js}$ denotes the nodes on the path from $v_j$ to server. Without loss of generality, we assume that the server is attached to the node located at the middle of the network, as changing the location of the server does not affect the scaling laws. Using the protocol model and according to \cite{book06:Xue},  the transport capacity in such network is upper bounded by $\Theta(W\sqrt{n})$. This is the model studied in \ref{thm:01} and the first two scenarios of Theorem \ref{thm:02}.

\subsubsection{Random Network}

The next network studied in Theorem \ref{thm:02} is a more general network model where the nodes are randomly distributed over a unit square area according to a uniform distribution. We use the same model used in \cite{book06:Xue} (section $5$) and divide the network area into square cells each with side-length proportional to the transmission range $r(n)$, which is selected to be at least in the order of $\sqrt{\frac{\log n}{n}}$ to guarantee the connectivity of the network \cite{Applied97:Penrose}. According to the protocol model \cite{book06:Xue}, if the cells are far enough they can transmit data at the same time with no interference; we assume that there are $M^2$ non-interfering groups which take turn to transmit at the corresponding time-slot in a round robin fashion. Again, without loss of generality the server is assumed to be located at the middle of the network. In this model the maximum number of simultaneous feasible transmissions  will be in the order of $\frac{1}{r^2(n)}$ as each transmission consumes an area proportional to $r^2(n)$.

All other assumptions are similar to the grid network.

\subsection{Content Discovery Algorithm}

\subsubsection{Path-wise Discovery}

To discover the location of the desired content, the request is sent through the shortest path toward the server containing the requested content. If an intermediate node has the data in its local cache, it does not forward the request toward the server anymore and the requester will start downloading from the discovered cache. Otherwise, the request will go all the way toward the server and the content is obtained from the main source. In case of the random network when a node needs a piece of information, it will send a request to its neighbors toward the server, i.e. the nodes in the same cell and one adjacent cell in the path toward the server, if any copy of the data is found it will be downloaded. If not, just one node in the adjacent cell will forward the request to the next cell toward the server.
	
\subsubsection{Expanding Ring Search}
In this algorithm the request for the information is sent to all the nodes in the transmission range of the requester. If a node receiving the request contains the required data in its local cache, it notifies the requester and then downloading from the discovered cache is started. Otherwise, all the nodes that receive the request will broadcast the request to their own neighbors. This process continues until the content is discovered in a cache and the downloading follows after that. This will return the nearest copy from the requester.

\subsection{Content Distribution in Steady-State}
\label{SSanalysis}

\begin{figure}[http]
    \center
      \includegraphics[scale=0.25,angle=0]{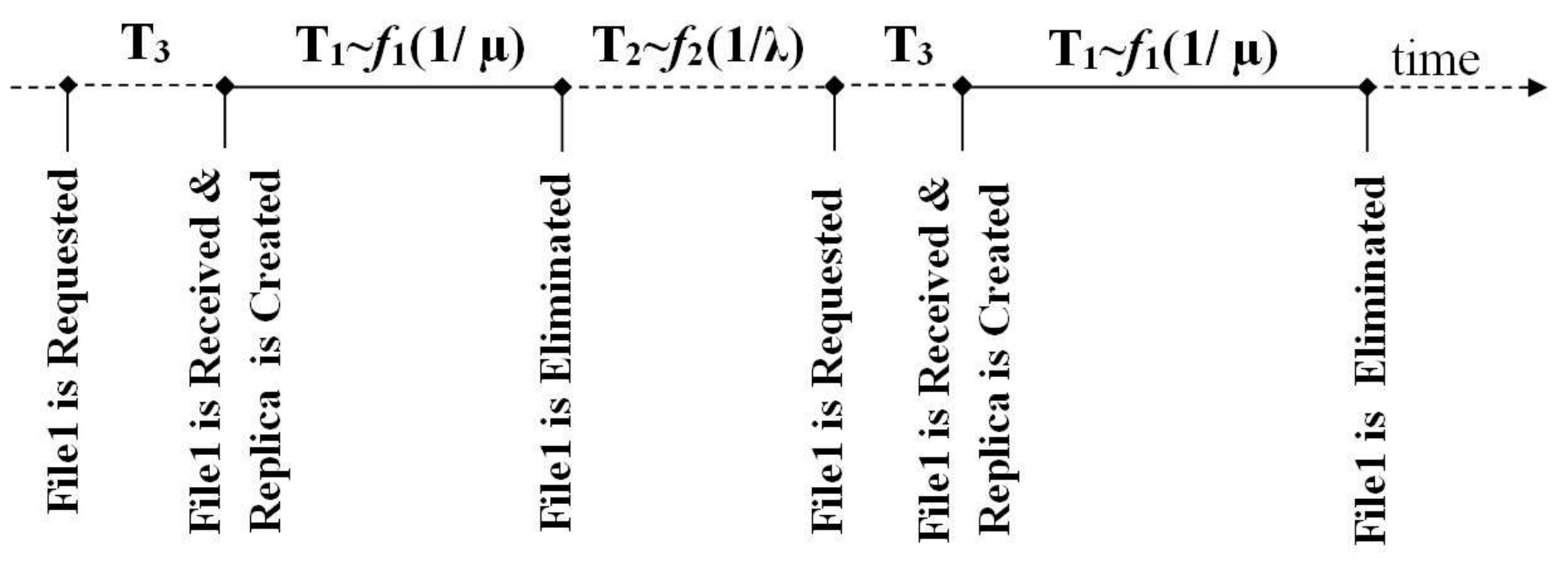}\\
      \caption{Data access process time diagram in a cache network}
    \label{fig:Tdgrm1}
\end{figure}

The time diagram of data access process in a cache is illustrated in Figure \ref{fig:Tdgrm1}.
When a query for content $f_i$ is initiated, the content is available at the requester's cache after a wait time ($T_3$) which is a function of the distance between the user and the data source (server or an intermediate cache), the data size, and the download speed. An expiration timer will be set upon receiving the data, and this data will be finally dropped after a holding time ($T_1$) with distribution $\mathfrak{f}_{1_i}$ and mean $1/\mu_i$. During this time, the cached data can be shared with the other users if needed. The same user may re-issue a query for that data after some random time ($T_2$) with distribution $\mathfrak{f}_{2_i}$ and mean $1/\lambda_i$. Note that a node will send out a request for a content only if it does not have it in its local cache, otherwise, its request will be served locally and no request is sent to the other nodes. The solid lines in this diagram denote the portions of time that the data is available at local cache.

In this work we assume identical content sizes $B_i=B$, and assume all the contents have the same popularity leading to similar request rates $\lambda_i=\lambda$, and the same holding times $\mu_i=\mu$. As the requests for different contents are supposed to be independent and holding times are set for each content independent of the others, we can do the calculations for one single content. If the total number of contents is not a function of the network size, this will not change the capacity order.
Suppose that $B$ is much larger than the request packet size, so we ignore the overhead of the discovery phase in our calculations. Furthermore, if the information sizes are the same and the download rates are also the same, the download time will be a function of the number of hops ($h$) between the source and the customer; $T_3=Bh/\gamma$. In the steady-state analysis, we ignore this constant time.

The average portion of time that each node contains a content in its local cache is
\begin{eqnarray}
\rho(n)=\frac{1/\mu}{1/\mu+1/\lambda}=\frac{\lambda}{\lambda+\mu},
\end{eqnarray}
which is the average probability that a node contains the data at steady-state. $\lambda$ is the rate of requests for a data from each user in case of the data not being  available, and $\mu$ is the rate of the data being expunged from the cache. Both these parameters are strongly dependent on the total number of users, or the topology and configuration of the network or the cache characteristics like size and replacement policy.

\section{Theorem Statements and Proofs}
\label{sec:theorems}

\begin{theorem} \label{thm:01}
Consider a grid wireless network consisting of $n$ nodes. Each node can transmit over a common wireless channel, with bandwidth $W$ bits per second, shared by all nodes. Assume that there is a server which contains all the information. Without loss of generality we assume that this server is located in the middle of the network. Each node contains some information in its local cache. Assume that the probability of the information being in all the caches with the same distance ($j$ hops) from the server is the same ($\rho_j(n)$). The maximum achievable throughput capacity order\footnote{$f(n)=O(g(n))$ or $f(n)\preceq g(n)$ if $sup_n(f(n)/g(n))<\infty$. $f(n)=\Omega(g(n))$ or $f(n) \succeq g(n)$ if $g(n)=O(f(n))$. $f(n)=\Theta(g(n))$ or $f(n)\equiv g(n)$ if both $f(n)=O(g(n))$ and $f(n)=\Omega(g(n))$. $f(n)=o(g(n))$ or $f(n) \prec g(n)$ if $f(n)/g(n)\rightarrow 0$. $f(n)=\omega(g(n))$ or $f(n) \succ g(n)$ if $g(n)/f(n)\rightarrow 0$.} ($\gamma_{max}$) in such network when the nodes use the nearest copy of the required content on the shortest path toward the server is given by

\begin{eqnarray}
			\gamma_{max}\equiv \frac{W\sqrt{n}}{\sum_{i=1}^{\sqrt{n}} i \sum_{j=0}^{i-1} (i-j)\rho_j(n)\prod_{k=j+1}^i(1-\rho_k(n))}, \nonumber
\end{eqnarray}
	
where $\rho_0(n)=1$, which means that the server always contains the information.

\end{theorem}

\begin{IEEEproof}
A request initiated by a user $v_i$ in $i$-hop distance from the server (located in level $i=1,..,\sqrt{n}$) is served by cache $u_j$ located in level $j,\ 1\leq j\leq i$ on the shortest path from $v_i$ to the server if no caches before $u_j$, including $v_i$, on this path contains the required information, and $u_j$ contains it. This request is served by the server if no copy of it is available on the path. Assuming that the availability of the information in each cache is independent of the contents in the other caches, this probability denoted by $P_{i,j}$ is given by

\begin{eqnarray}
P_{i,j}=(1-\rho_i(n))(1-\rho_{i-1}(n))...(1-\rho_{j+1}(n))\rho_j(n)
\end{eqnarray}

where $\rho_j(n)$ is the probability of the information being available in a cache in level $j, 1 \leq j \leq \sqrt{n}$, and $j=0$ shows the server and $\rho_0(n)=1$. Thus a content requested by $v_i$ is traveling $i-j$ hops with probability $P_{i,j}$. There are $4i$ nodes in level $i$ so the average number of hops ($E[h]$) traveled by each piece of data from the serving cache (or the original server) to the requester is

\begin{equation}
E[h]=\frac{1}{n}\sum_{i=1}^{\sqrt{n}} 4i \sum_{j=0}^{i-1} (i-j)P_{i,j}, \nonumber 
\end{equation}
\begin{equation}
=\frac{1}{n}\sum_{i=1}^{\sqrt{n}} 4i \sum_{j=0}^{i-1} (i-j)(1-\rho_i(n))...(1-\rho_{j+1}(n))\rho_j(n).  \label{eq:barh} 
\end{equation}

Assume that each user is receiving data with rate $\gamma$. The transport capacity in this network, which equals to $n\gamma E[h]$, is upper bounded by $\Theta(W\sqrt{n})$. So $\gamma_{max}=\Theta(\frac{W}{E[h]\sqrt{n}})$ and the Theorem is proved.
\end{IEEEproof}

\begin{theorem} \label{thm:02}
	Consider a wireless network consisting of $n$ nodes, with each node containing the information in its local cache with common probability $\rho(n)\nrightarrow 1$.\footnote{Note that for $\rho (n) \rightarrow 1$, the request is served locally and no data is transferred between the nodes.} Assume that the request process and cache look up time in each node is not a function of the number of nodes, then
	
	\begin{itemize}
		\item Scenario $\romannumeral 1$- If the nodes are located on a grid and search for the contents just on the shortest path toward the server, the average delay order is
		
		\begin{eqnarray}
			\left\{\begin{array}{ll}
					\Theta(\sqrt{n})& ,if\ \rho(n)\preceq \frac{1}{\sqrt{n}} \\
					\Theta(\frac{1}{\rho(n)})& ,if\ \rho(n)\succeq \frac{1}{\sqrt{n}} \nonumber
		\end{array}\right .
		\end{eqnarray}
		
		\item Scenario $\romannumeral 2$- If the nodes are located on a grid and use ring expansion as their content search algorithm, the average delay order is
		\begin{eqnarray}
			\left\{\begin{array}{ll}
					\Theta(\sqrt{n})& ,if\ \rho(n)\preceq \frac{1}{n} \\
					\Theta(\frac{1}{\sqrt{\rho(n)}})& ,if\ \rho(n)\succeq \frac{1}{n} \nonumber
		\end{array}\right .
		\end{eqnarray}		
		
		\item Scenario $\romannumeral 3$- If the nodes are randomly distributed over a unit square area and use path-wise content discovery algorithm, the average delay order is 	
		
		\begin{eqnarray}
			\left\{\begin{array}{ll}
					\Theta(\sqrt{\frac{n}{\log n}})& ,if\ \rho(n)\preceq \frac{1}{\sqrt{n\log n}}  \\
					\Theta(\frac{1}{\rho(n)\log n})& ,if\ \frac{1}{\sqrt{n\log n}} \preceq \rho(n) \preceq \frac{1}{\log n} \\
					\Theta(1)& ,if\  \rho(n)\succeq \frac{1}{\log n}\nonumber
		\end{array}\right .
		\end{eqnarray}
		\end{itemize}
\end{theorem}

Here we prove Theorem \ref{thm:02} by utilizing some Lemmas.

\begin{lemma} \label{lem:01}
	Consider the wireless networks described in Theorem \ref{thm:02}. The average number of hops between the customer and the serving node (a cache or original server) is 
	
		\begin{itemize}
		\item Scenario $\romannumeral 1$
			\begin{eqnarray}
		&E[h]\equiv \frac{1}{n}\sum_{i=1}^{\sqrt{n}} i^2(1-\rho(n))^i& \nonumber \\
		&+ \frac{\rho(n)}{n}\sum_{i=1}^{\sqrt{n}}i\sum_{k=1}^{i-1}k(1-\rho(n))^k& \label{eq:EXi}
	\end{eqnarray}

		\item Scenario $\romannumeral 2$
		\begin{equation}
		E[h]\equiv \frac{1}{n}\sum_{i=1}^{\sqrt{n}} i^2(1-\rho(n))^{2i^2-2i+1} \nonumber 
		\end{equation}
		\begin{equation}
		+ \frac{1}{n}\sum_{i=2}^{\sqrt{n}}i\sum_{k=1}^{i-1}k(1-\rho(n))^{2k^2-2k+1}(1-(1-\rho(n))^{4k}) \label{eq:EXii}
	  \end{equation}
		
		\item Scenario $\romannumeral 3$
		\begin{equation}
		E[h]\equiv \frac{\log n}{n}\sum_{i=2}^{\sqrt{\frac{n}{\log n}}} i^2(1-\rho(n))^{i\log n} \nonumber 
		\end{equation}
		\begin{equation}
		+ \frac{\log n(1-(1-\rho(n))^{\log n})}{n}\sum_{i=2}^{\sqrt{\frac{n}{\log n}}}i\sum_{k=1}^{i-1}k(1-\rho(n))^{k\log n} \label{eq:EXiii}
	\end{equation}
	
		\end{itemize}
	\end{lemma}
	
\begin{IEEEproof}
Let $h$, $d_{sr}$, and $d_{max}$ denote the number of hops between the customer and the serving node (cache or original server), the number of hops between the customer and the original server, and the maximum value of $d_{sr}$, respectively. The average number of hops between the customer and the serving node ($E[h]$) is given by

\begin{eqnarray}
E[h]=\sum_{i=1}^{d_{max}} E[h|d_{sr}=i]Pr(d_{sr}=i) \label{eq:hbar}
\end{eqnarray}

Scenario $\romannumeral 1$- This case can be considered as a special case of the network studied in theorem \ref{thm:01}, where $\rho_i(n)$ is the same for all $i$. Thus we can drop the index $i$ and let $\rho(n)$ denote the common value of this probability. Using equation \ref{eq:barh} we will have

\begin{equation}
E[h]\equiv \frac{4}{n}\sum_{i=1}^{\sqrt{n}} i \{i(1-\rho(n))^i+\sum_{j=1}^{i-1} (i-j)(1-\rho(n))^{i-j}\rho(n)\} 
\end{equation}

The constant factor $4$ does not have any affect on the scaling order, so it can be dropped. Using variable $k=i-j$ then proves the Lemma.

\begin{equation}
E[h]\equiv\frac{1}{n} [ \sum_{i=1}^{\sqrt{n}} i^2(1-\rho(n))^i +\sum_{i=1}^{\sqrt{n}}i\sum_{k=1}^{i-1} k(1-\rho(n))^k\rho(n)] 
\end{equation}

Scenario $\romannumeral 2$ - $d_{max}$ in this network is $\Theta(\sqrt{n})$, and there are $4i$ nodes at distance of $i$ hops from the original server. 
\begin{equation}
Pr(d_{sr}=i)\equiv\frac{i}{n}
\end{equation}

Each customer may have the required item in its local cache with probability $\rho(n)$. If the requester is one hop away from the original server, it gets the required item from the server with probability $1-\rho(n)$. The customers at two hops distance from the server ($8$ such customers) download the required item from the original server (traveling $h=2$ hops) if no cache in a diamond of two hops diagonals contains it (probability $(1-\rho(n))^2$), and gets it from a cache at distance one hop if one of those caches has the item (probability $(1-\rho(n))(1-(1-\rho(n))^4)$). Using similar reasoning, the customers at distance $i$ from the server get the item from the server (distance $h=i$ hops) with probability $(1-\rho(n))^{1+4(1+2+...+(i-1))}=(1-\rho(n))^{2i^2-2i+1}$, and from a cache at distance $h=k<i$ with probability $(1-\rho(n))^{2k^2-2k+1}(1-(1-\rho(n))^{4k})$ as there are $4k$ nodes at distance of $k$ hops. Therefore, using equations (\ref{eq:hbar}) and (\ref{eq:barh})

\begin{equation}
E[h]\equiv \frac{1}{n}\sum_{i=2}^{\sqrt{n}} i\sum_{k=1}^{i-1} k(1-(1-\rho(n))^{4k})(1-\rho(n))^{2k^2-2k+1} \nonumber
\end{equation}
\begin{equation}
 +  \frac{1}{n}\sum_{i=1}^{\sqrt{n}} i^2(1-\rho(n))^{2i^2-2i+1}
\end{equation}

Scenario $\romannumeral 3$ -  Each hop is one cell containing $\Theta(\log n)$ caches. $d_{max}$ in this network is of the order of $\sqrt{\frac{n}{\log n}}$ and
$\Pr(d_{sr}=i) \equiv \frac{i \log n}{n}$.

Each customer may have the required item in its local cache with probability $\rho(n)$.  If the requester is one hop away from the original server ($4\Theta(\log n)$ nodes), it gets the required item from the server with probability $1-\rho(n)$. The customers at two hops distance from the server ($8\Theta(\log n)$ such customers) download the required item from the original server (traveling $h=2$ hops) if no cache in the cell at one hop distance contains it (probability $(1-\rho(n))^{2\log n}$), and gets it from a cache at distance one hop if one of those caches has the item (probability $(1-\rho(n))(1-(1-\rho(n))^{2\log n})$). Using similar reasoning the customers at distance $i$ from the server get the item from the server (distance $h=i$ hops) with probability $(1-\rho(n))^{i\log n}$, and from a cache at distance $h=k<i$ with probability $(1-\rho(n))^{k\log n}(1-(1-\rho(n))^{\log n})$. Therefore, according to equation (\ref{eq:hbar})

\begin{equation}
		E[h]\equiv \frac{\log n}{n}(1-\rho(n))+\frac{\log n}{n}\sum_{i=2}^{\sqrt{\frac{n}{\log n}}} i^2(1-\rho(n))^{i\log n} \nonumber 
\end{equation}
\begin{equation}
		+ \frac{\log n(1-(1-\rho(n))^{\log n})}{n}\sum_{i=2}^{\sqrt{\frac{n}{\log n}}}i\sum_{k=1}^{i-1}k(1-\rho(n))^{k\log n}.
	\end{equation}

	Noting that $\frac{\log n}{n}(1-\rho(n))$ is always less than one, and tends to zero for sufficiently large $n$, the Lemma is proved.
	\end{IEEEproof}
	
	\begin{lemma}\label{lem:02}
	Consider the wireless networks described in Theorem \ref{thm:02}. For sufficiently large networks, the average number of hops between the customer and the serving node (a cache or the original server) is 
	
	\begin{itemize}
		\item Scenario $\romannumeral 1$
			\begin{eqnarray}
		&E[h]\equiv\left\{\begin{array}{ll}
					\sqrt{n}& \ \ \ \rho(n) \preceq \frac{1}{\sqrt{n}} \\
					\frac{1}{\rho(n)}& \ \ \ \rho(n) \succeq \frac{1}{\sqrt{n}}  \label{eq:hi} \\
		\end{array}\right .
	\end{eqnarray}

		\item Scenario $\romannumeral 2$
		\begin{eqnarray}
		&E[h]\equiv\left\{\begin{array}{ll}
					\sqrt{n}& \ \ \ \rho(n) \preceq \frac{1}{n} \\
					\frac{1}{\sqrt{\rho(n)}}& \ \ \ \rho(n) \succeq \frac{1}{n} \label{eq:hii} \\
		\end{array}\right .
	\end{eqnarray}
	
		\item Scenario $\romannumeral 3$
		\begin{eqnarray}
		&E[h]\equiv\left\{\begin{array}{ll}
					\sqrt{\frac{n}{\log n}}& \ \ \ \rho(n) \preceq \frac{1}{\sqrt{n\log n}} \\
					\frac{1}{\rho(n) \log n}& \ \ \ \frac{1}{\sqrt{n\log n}} \preceq \rho(n) \preceq \frac{1}{\log n} \\
					1& \ \ \ \rho(n) \succeq \frac{1}{\log n} \label{eq:hiii} \\
		\end{array}\right .
	\end{eqnarray}
	
		\end{itemize}
\end{lemma}

\begin{IEEEproof}
To prove this Lemma we use the following equation which is true for every $N$ and $x$.
\begin{eqnarray}
\lim_{N\rightarrow \infty} (1-x)^N=\left\{\begin{array}{ll}
					1& \ \ \  x=o(\frac{1}{N}) \\
					e^{-xN}& \ \ \  x=\Theta(\frac{1}{N}) \\
					0& \ \ \ x=\omega(\frac{1}{N})
		\end{array}\right . \label{eq:e_xn}
\end{eqnarray}

Scenario $\romannumeral 1$ - Let's define 
\begin{eqnarray}
E_s^i&=&\frac{1}{n}\sum_{i=1}^{\sqrt{n}} i^2(1-\rho(n))^i,  \\
E_c^i&=&\frac{\rho(n)}{n}\sum_{i=1}^{\sqrt{n}}i\sum_{k=1}^{i-1}k(1-\rho(n))^k.
\end{eqnarray}

Thus equation (\ref{eq:EXi}) is written as $E[h]=E_s^i+E_c^i$.
First we investigate the value of $E_s^i$ for different ranges of $\rho(n)$. The summation for $E_s^i$ can be decomposed into two  summations.

\begin{eqnarray}
E_s^i&\equiv& \frac{1}{n}(\sum_{i\prec \sqrt{n}}i^2(1-\rho(n))^i + \sum_{i\equiv \sqrt{n}}i^2(1-\rho(n))^i) 
\end{eqnarray}

Assume $\rho(n)\equiv \frac{1}{\sqrt{n}}$, then using first and second region of equation (\ref{eq:e_xn}) we have
\begin{eqnarray}
E_s^i\equiv \frac{1}{n}(\sum_{i\prec \sqrt{n}}i^2 + \sum_{i\equiv \sqrt{n}}i^2) \equiv \frac{n^{3/2}}{n} \equiv \sqrt{n}.
\end{eqnarray}

Moreover it can easily be seen that $E_s^i$ is a decreasing function of $\rho(n)$, so for $\rho(n)$ with order less than $\frac{1}{\sqrt{n}}$ it is more than $\sqrt{n}$. Since $d_{max}=\sqrt{n}$, we can say $E_s^i\equiv \sqrt{n}$ for $\rho(n) \preceq \frac{1}{\sqrt{n}}$.

Now we expand the summation to obtain 
\begin{equation}
E_s^i =\frac{(1-\rho(n))(2-\rho(n))}{n\rho^3(n)} \nonumber
\end{equation}
\begin{equation}
-\frac{(1-\rho(n))^{\sqrt{n}+1}}{n\rho^3(n)} \times \nonumber 
\end{equation}
\begin{equation}
(n(1-\rho(n))^2-(1-\rho(n))(2n+2\sqrt{n}-1)+(\sqrt{n}+1)^2)
\end{equation}

when $\rho(n) \succ \frac{1}{\sqrt{n}}$ then using third region in equation \ref{eq:e_xn}, $(1-\rho(n))^{\sqrt{n}+1}$ is going to zero exponentially, so $n(1-\rho(n))^{\sqrt{n}+1}\rightarrow 0$. Thus, $E_s^i \equiv \frac{1}{n\rho^3(n)}$.

\begin{eqnarray}
E_s^i \equiv \left\{\begin{array}{ll}
					\sqrt{n}& \ \ \  \rho(n)\preceq \frac{1}{\sqrt{n}} \\
					\frac{1}{n\rho^3(n)}& \ \ \  \rho(n)\succ \frac{1}{\sqrt{n}}
			\end{array}\right . \label{eq:Esi}
\end{eqnarray}

According to equation (\ref{eq:Esi}) and since $E[h]=E_s^i+E_c^i$, when $E_s^i\equiv \sqrt{n}$ (for $\rho(n)\preceq \frac{1}{\sqrt{n}}$) which is the maximum possible order for $E[h]$, then adding $E_s^i$ to $E[h]$ cannot increase its order beyond the maximum possible value.
Now to derive the order of $E[h]$ for other values of $\rho(n)$, we decompose the equation of $E_c^i$ to the following summations and investigate their behaviors when $\rho(n)\succ \frac{1}{\sqrt{n}}$.

\begin{eqnarray}
E_c^i&=& E_c^{i1}+E_c^{i2} \nonumber \\
E_c^{i1}&=&\frac{1}{n}\sum_{i\equiv \sqrt{n}}i\sum_{k=1}^{i-1}k\rho(n)(1-\rho(n))^k \nonumber \\
E_c^{i2}&=&\frac{1}{n}\sum_{i\prec \sqrt{n}}i\sum_{k=1}^{i-1}k\rho(n)(1-\rho(n))^k 
\end{eqnarray}

The number of $i\equiv \sqrt{n}$ is in the order of $\Theta(1)$. Therefore using the following series $\sum_{x=1}^{n} x a^x$ $= \frac{a^{n+1}(n a - n - 1)+a}{(a-1)^2}$, we have 
\begin{eqnarray}
E_c^{i1}&\equiv& \frac{1}{\sqrt{n}}\sum_{k=1}^{\sqrt{n}}k\rho(n)(1-\rho(n))^k, \nonumber \\
&\equiv& \frac{1-\rho(n)}{\rho(n)\sqrt{n}}(1-(1-\rho(n))^{\sqrt{n}}(1+\rho(n)\sqrt{n})), \nonumber 
\end{eqnarray}
which is equivalent to $\frac{1}{\rho(n)\sqrt{n}}$ when $\rho(n)\succ \frac{1}{\sqrt{n}}$.

Utilizing the same series, the first summation in $E_c^{i2}$ is in the order of $\sqrt{n}$. Hence we arrive at
\begin{equation}
E_c^{i2}\equiv \frac{1-\rho(n)}{\rho(n)n}\sum_{i\prec \sqrt{n}} i[1-\{1-\rho(n)+\rho(n)i\}(1-\rho(n))^{i-1}] \nonumber 
\end{equation}
\begin{equation}
\equiv \frac{1-\rho(n)}{\rho(n)} \times \nonumber
\end{equation}
\begin{equation}
- \frac{1}{n}\sum_{i\prec \sqrt{n}} i(1-\rho(n))^i \nonumber 
\end{equation}
\begin{equation}
- \frac{1}{n}\sum_{i\prec \sqrt{n}} i^2\rho(n)(1-\rho(n))^{i-1} \nonumber 
\end{equation}
\begin{equation}
\equiv \frac{1-\rho(n)}{\rho(n)}-\frac{(1-\rho(n))^2}{\rho^3(n)n}-\frac{1}{\rho^3(n)n} \nonumber 
\end{equation}
\begin{equation}
\equiv \frac{1}{\rho(n)}
\end{equation}

Since  $\rho(n)\succ \frac{1}{\sqrt{n}}$, $E_c^{i2}$ is the dominant factor in $E_c^i$, and also it is dominant factor in $E[h]$. Thus,

\begin{eqnarray}
E[h]\equiv \left\{\begin{array}{ll}
					E_s^i \equiv \sqrt{n}& \ \ \  \rho(n)\preceq \frac{1}{\sqrt{n}} \\
					E_c^{i2}\equiv \frac{1}{ \sqrt{\rho(n)}}& \ \ \  \rho(n)\succ \frac{1}{\sqrt{n}}
			\end{array}\right . 
\end{eqnarray}

Scenario $\romannumeral 2$ - Let's define 
\begin{equation}
E_s^{ii}= \frac{1}{n}\sum_{i=1}^{\sqrt{n}} i^2(1-\rho(n))^{2i^2-2i+1}, \nonumber 
\end{equation}
\begin{equation}
E_c^{ii} \frac{1}{n}\sum_{i=2}^{\sqrt{n}}i\sum_{k=1}^{i-1}k(1-\rho(n))^{2k^2-2k+1}(1-(1-\rho(n))^{4k}), \nonumber 
\end{equation}
\begin{equation}
E[h]= E_s^{ii}+E_c^{ii}.
\end{equation}

Assume that $\rho(n)\equiv \frac{1}{n}$, then 
\begin{eqnarray}
E_s^{ii}&\equiv& \frac{1}{n}\sum_{i=1}^{\sqrt{n}} i^2(1-\frac{1}{n})^{2i^2-2i+1}, \nonumber \\
&\equiv& \frac{1}{n}\sum_{i=1}^{\sqrt{n}} i^2 \equiv \sqrt{n}.
\end{eqnarray}
Since $E_s^{ii}$ is increasing when  $\rho(n)$ is decreasing and its maximum possible order is $\sqrt{n}$, then $E_s^{ii}\equiv \sqrt{n}$ for all $\rho(n) \preceq \frac{1}{n}$.

For $\rho(n) \succ \frac{1}{n}$, we approximate the summation with the integral.

\begin{equation}
E_s^{ii}\equiv \frac{1}{n}\int_{v=1}^{\sqrt{n}} v^2(1-\rho(n))^{2v^2-2v+1} \nonumber 
\end{equation}
\begin{eqnarray}
&\equiv \frac{(1-\log(1-\rho(n)))\sqrt{2\pi(1-\rho(n))} erf(\frac{(2v-1)\sqrt{-\log(1-\rho(n))}}{\sqrt{2}})}{n\log^{3/2}(1-\rho(n))}&
 \nonumber \\
&+  \frac{-2\sqrt{-\log(1-\rho(n))}(2v+1)(1-\rho(n))^{2v^2-2v+1}}{n\log^{3/2}(1-\rho(n))}|_{v=1}^{\sqrt{n}}&
\end{eqnarray}

where $erf$ is the error function which is always limited by $[-1,1]$ and is zero at zero.
If $\rho(n) \rightarrow 1$, then it is obvious that $E_s^{ii}\rightarrow 0$. For other values of $\rho(n) \succ \frac{1}{n}$ we use the third approximation in equation \eqref{eq:e_xn}, and also\footnote{This is true when $\rho(n)$ tends to zero while n approaches infinity.} $-\log(1-\rho(n))\equiv \rho(n)$ for $\rho(n)\rightarrow 0$ and $-\log(1-\rho(n))\equiv 1$ for $\rho(n)\nrightarrow 0$ to obtain

\begin{eqnarray}
E_s^{ii} \equiv \left\{\begin{array}{ll}
					\sqrt{n}& \ \ \  \rho(n)\preceq \frac{1}{n} \\
					\frac{1}{n\rho^{3/2}(n)}& \ \ \  \rho(n)\succ \frac{1}{n}
			\end{array}\right . \label{eq:Esii}
\end{eqnarray}
Since for $\rho(n)\preceq \frac{1}{n}$ the $E_s^{ii}$ reaches the maximum $E[h]$, therefore $E_c^{ii}$ cannot increase the scaling value of $E[h]$ anymore.
For $\rho \succ \frac{1}{n}$  we have
\begin{eqnarray}
E_c^{ii} \equiv \sqrt{\frac{1}{ \rho(n)}}. \label{eq:Ecii}
\end{eqnarray}

Thus it can easily be verified that

\begin{eqnarray}
E[h]\equiv \left\{\begin{array}{ll}
					E_s^{ii} \equiv \sqrt{n},& \ \ \  \rho(n)\preceq \frac{1}{n} \\
					E_c^{ii}\equiv \sqrt{\frac{1}{ \rho(n)}}.& \ \ \  \rho(n)\succ \frac{1}{n}
			\end{array}\right . 
\end{eqnarray}

Scenario $\romannumeral 3$ - Let's define

\begin{equation}\label{eq33}
E_s^{iii}= \frac{\log n}{n}\sum_{i=2}^{\sqrt{\frac{n}{\log n}}} i^2(1-\rho(n))^{i\log n} \nonumber 
\end{equation}
\begin{eqnarray}
&E_c^{iii}= & \nonumber \\
&\frac{\log (n)(1-(1-\rho(n))^{\log n})}{n}\sum_{i=2}^{\sqrt{\frac{n}{\log n}}}i\sum_{k=1}^{i-1}k(1-\rho(n))^{k\log n}& \nonumber 
\end{eqnarray}
\begin{equation}
E[h]\equiv E_s^{iii}+E_c^{iii}
	\end{equation}
	
	First we check the behavior of $E_s^{iii}$ when $\rho(n)\equiv \frac{1}{\sqrt{n\log n}}$. Using second region in equation \eqref{eq:e_xn} we will have $E_s^{iii}\equiv \sqrt{\frac{n}{\log n}}$. $E_s^{iii}$ is increasing when $\rho(n)$ is decreasing and the maximum possible value for the number of hops is $\sqrt{\frac{n}{\log n}}$, then $E_s^{iii} \equiv \sqrt{\frac{n}{\log n}}$ for all $\rho(n) \preceq \frac{1}{\sqrt{n\log n}}$.
	
	By approximating the summation with integral, we arrive at
	\begin{equation}
	E_s^{iii}\equiv \frac{\log n}{n}\int_2^{\sqrt{\frac{n}{\log n}}} v^2(1-\rho(n))^{v\log n}, \nonumber 
	\end{equation}
\begin{equation}
	\equiv \{\frac{\log (n)(1-\rho(n))^{v\log n}}{n\log^3 {(1-\rho(n))^{\log n}}} \times  \nonumber \\
	\end{equation}
	\begin{equation}
	(v^2\log^2 {(1-\rho(n))^{\log n}}-2v\log {(1-\rho(n))^{\log n}} +2)\}|_{v=2}^{\sqrt{\frac{n}{\log n}}}. 
	\end{equation}

	If $\frac{1}{\sqrt{n\log n}} \preceq \rho(n) \preceq \frac{1}{\log n}$, using equation \eqref{eq:e_xn} and the fact $\log {(1-\rho(n))^{\log n}}\equiv -\rho(n)\log n$, we will have
	\begin{eqnarray}
	E_s^{iii}&\equiv& \frac{1}{n\rho^3(n) \log^2 n}.
	\end{eqnarray}	
	When $\rho(n) \succeq \frac{1}{\log n}$, equation \eqref{eq33} tends to zero.
	\begin{eqnarray}
	E_s^{iii}&\equiv& \left\{\begin{array}{ll}
					\sqrt{\frac{n}{\log n}}& \ \ \  \rho(n)\preceq \frac{1}{\sqrt{n\log n}} \\
					\frac{1}{n\rho^3(n) \log^2 n}& \ \ \  \frac{1}{\sqrt{n\log n}} \preceq \rho(n)\preceq \frac{1}{\log n} \\
					0& \ \ \ \rho(n)\succeq \frac{1}{\log n}
			\end{array}\right . 
	\end{eqnarray}

	Using the previous approximations along with $1-(1-\rho(n))^{\log n}\equiv 1$ for $\rho(n)\succeq \frac{1}{\log n}$ and $\rho(n)\log n$ for $\rho(n)\preceq \frac{1}{\log n}$, we can approximate $E_c^{iii}$ as its dominant terms.
	
	\begin{eqnarray}
	E_c^{iii} \equiv \frac{1}{n\rho(n)}\sum_{i=2}^{\sqrt{\frac{n}{\log n}}} i \equiv \frac{1}{\rho(n)\log n} 
	\end{eqnarray}
	
	When $\rho(n)\succeq \frac{1}{\log n}$, the dominant term is $\Theta(1)$. Thus,
	
	\begin{equation}
	E[h]\equiv \left\{\begin{array}{ll}
					E_s^{iii} \equiv \sqrt{\frac{n}{\log n}}& \ \ \  \rho(n)\preceq \frac{1}{\sqrt{n\log n}} \\
					E_c^{iii}\equiv \frac{1}{\rho(n)\log n}& \ \ \  \frac{1}{\sqrt{n\log n}} \preceq \rho(n)\preceq \frac{1}{\log n} \\
					E_c^{iii}\equiv 1& \ \ \  \frac{1}{\log n} \preceq \rho(n)
			\end{array}\right . 
	\end{equation}
	
	It can be seen that for large enough $\rho(n)$ the average number of hops between the nearest content location and the customer is just $\Theta(1)$ hops. This is the result of having $log(n)$ caches in one hop distance for  every requester. Each one of these caches can be  a potential source for the content. When the network grows, this number will increase and if $\rho(n)$ is large enough ($\frac{1}{\log n} \preceq \rho(n)$) the probability that at least one of these nodes contain the required data will approach 1, i.e., $\lim_{n\rightarrow \infty}  (1-(1-\rho(n))^{\log n}) = 1$.
	
	\end{IEEEproof}
	
	Theorem \ref{thm:02} is now simply proved using the above Lemmas.

\begin{IEEEproof}
Assuming that the delay of the request process and cache look up in each node is not increasing when the network size (the number of nodes) increases, and the there is enough bandwidth to avoid congestion, then the delay of getting the data is directly proportional to the average number of hops between the serving node and the customer. Thus, the delay and the average number of hops the data is traveling to reach the customer are of the same order and \textit{Theorem \ref{thm:02}} is proved.
\end{IEEEproof}

\begin{theorem}\label{thm:03}
Consider the networks of Theorem \ref{thm:02}, and assume each node can transmit over a common wireless channel, with $W$ bits per second bandwidth, shared by all nodes. The maximum achievable throughput capacity order $\gamma_{max}$  in the three discussed scenarios are
	
	\begin{itemize}
		\item Scenario $\romannumeral 1$- 
		
		\begin{eqnarray}
			\left\{\begin{array}{ll}
					\Theta(\frac{W\rho(n)}{\sqrt{n}})& ,if\ \rho(n)\succeq \frac{1}{\sqrt{n}} \\
					\Theta(\frac{W}{n})& ,if\ \rho(n)\preceq \frac{1}{\sqrt{n}} \nonumber
		\end{array}\right .
		\end{eqnarray}
		
		\item Scenario $\romannumeral 2$- 
		\begin{eqnarray}
			\left\{\begin{array}{ll}
					\Theta(W\sqrt{\frac{\rho(n)}{n}})& ,if\ \rho(n)\succeq \frac{1}{n} \\
					\Theta(\frac{W}{n})& ,if\ \rho(n)\preceq \frac{1}{n} \nonumber
		\end{array}\right .
		\end{eqnarray}		
		
		\item Scenario $\romannumeral 3$-	
		
		\begin{eqnarray}
			\left\{\begin{array}{ll}
					\Theta(\frac{W}{\log n})& ,if\ \rho(n)\succeq \frac{1}{\log n} \\
					\Theta(\rho^2(n)\log n W)& ,if\ \frac{1}{\sqrt{n\log n}} \preceq \rho(n) \preceq \frac{1}{\log n} \\
					\Theta(\frac{W}{n})& ,if\ \rho(n)\preceq \frac{1}{\sqrt{n\log n}} \nonumber
		\end{array}\right .
		\end{eqnarray}
		\end{itemize}
\end{theorem}

To prove Theorem \ref{thm:03} we use Lemma \ref{lem:02}, and the following two Lemmas.

\begin{lemma}\label{lem:03}
	Consider the wireless networks described in Theorem \ref{thm:02}. In order not to have interference, the maximum throughput capacity is upper limited by  
	
	\begin{itemize}
		\item Scenario $\romannumeral 1$- 
		
		\begin{eqnarray}
			\left\{\begin{array}{ll}
					\Theta(\frac{W\rho(n)}{\sqrt{n}})& ,if\ \rho(n)\succeq \frac{1}{\sqrt{n}} \\
					\Theta(\frac{W}{n})& ,if\ \rho(n)\preceq \frac{1}{\sqrt{n}} \nonumber
		\end{array}\right .
		\end{eqnarray}
		
		\item Scenario $\romannumeral 2$- 
		\begin{eqnarray}
			\left\{\begin{array}{ll}
					\Theta(W\sqrt{\frac{\rho(n)}{n}})& ,if\ \rho(n)\succeq \frac{1}{n} \\
					\Theta(\frac{W}{n})& ,if\ \rho(n)\preceq \frac{1}{n} \nonumber
		\end{array}\right .
		\end{eqnarray}	
		
		\item Scenario $\romannumeral 3$-	
		
		\begin{eqnarray}
			\left\{\begin{array}{ll}
					\Theta(\frac{W}{\log n})& ,if\ \rho(n)\succeq \frac{1}{\log n} \\
					\Theta(\rho W)& ,if\ \frac{1}{\sqrt{n\log n}} \preceq \rho(n) \preceq \frac{1}{\log n} \\
					\Theta(\frac{W}{\sqrt{n \log n}})& ,if\ \rho(n)\preceq \frac{1}{\sqrt{n\log n}} \nonumber
		\end{array}\right .
		\end{eqnarray}
		\end{itemize}
\end{lemma}	
		
\begin{IEEEproof}
Assume that each content is retrieved with rate $\gamma$ bits/sec. The traffic generated because of one download from a cache (or server) at average distance of $E[h]$ hops from the requester node is $\gamma E[h]$. The total number of requests for a content in the network at any given time is limited by the number of nodes $n$. Thus the maximum total bandwidth needed to accomplish these downloads will be $nE[h]\gamma$, which is upper limited by ($\Theta(W\sqrt{n})$) in scenarios $\romannumeral 1$, $\romannumeral 2$, and ($\Theta(\frac{W}{r^2(n)})=\Theta(\frac{Wn}{\log n})$) in scenario $\romannumeral 3$. Thus,

\begin{eqnarray}
nE[h]\gamma &\preceq&  W\sqrt{n} \nonumber \\
\gamma_{max} &\equiv& \frac{W}{\sqrt{n}E[h]} 
\end{eqnarray}

in scenarios $\romannumeral 1$, $\romannumeral 2$, and

\begin{eqnarray}
nE[h]\gamma &\preceq&  \frac{Wn}{\log n} \nonumber \\
\gamma_{max} &\equiv& \frac{W}{\log nE[h]} 
\end{eqnarray}

in scenarios $\romannumeral 3$. Therefore the maximum download rate is easily derived using the results of Lemma \ref{lem:02}.
\end{IEEEproof}

In the previous Lemma, the maximum throughput capacity in a wireless network utilizing caches has been calculated such that no interference occurs. Now it is important to verify if this throughput can be supported by each node (cell), i.e. the traffic carried by each node (cell) is not more than what it can support ($\Theta(1)$).

\begin{lemma} \label{lem:04}

	The throughput capacities of Lemma \ref{lem:03} are supported for all values of $\rho(n)$ in grid topology. The random network can support the obtained throughput capacities just when $\rho(n) \succeq \frac{1}{\log n}$. For smaller values of $\rho(n)$ the maximum supportable throughput capacities are as follows.
	\begin{eqnarray}
	\gamma_{max} \equiv \left\{\begin{array}{ll}
					\frac{1}{n}& \ \ \  \rho(n)\preceq \frac{1}{\sqrt{n\log n}} \\
					\rho^2(n)\log n& \ \ \  \frac{1}{\sqrt{n\log n}} \preceq \rho(n)\prec \frac{1}{\log n} 
			\end{array}\right . 
	\end{eqnarray}
\end{lemma}

\begin{IEEEproof}
Each link between two nodes in scenarios $\romannumeral 1$ and $\romannumeral 2$, or two cells in scenario $\romannumeral 3$ can carry at most $\Theta(1)$ bits per second. Here we calculate the maximum traffic passing through a link considering the throughput capacities derived in previous Theorems, and check if any link can be a bottleneck.

Scenario $\romannumeral 1$-  Each one of the four links connected to the server will carry all the traffic related to the items not found in the on-path caches. Thus, the total traffic carried by each of those links is
$\sum_{i=1}^{\sqrt{n}}\gamma i (1-\rho(n))^i$.

When $\rho(n) \preceq \frac{1}{\sqrt{n}}$, we have $(1-\rho(n))^i \equiv 1$ for all $i\leq \sqrt{n}$. So this traffic is equal to
\begin{equation}
\sum_{i=1}^{\sqrt{n}}\gamma i \equiv n\gamma \preceq n\gamma_{max} \equiv 1.
\end{equation}

When $\rho(n) \succeq \frac{1}{\sqrt{n}}$, using equation \ref{eq:e_xn} the above summation can be written as

\begin{eqnarray}
\gamma \{\frac{(1-\rho(n))^{\sqrt{n}}(\sqrt{n}\log (1-\rho(n))-1)}{\log^2 (1-\rho(n))} \nonumber \\
- \frac{(1-\rho(n))(\log (1-\rho(n))-1)}{\log^2 (1-\rho(n))} \} \nonumber 
\end{eqnarray}
\begin{eqnarray}
&\equiv& \gamma \frac{(1-\rho(n))(-\log (1-\rho(n))+1)}{\log^2 (1-\rho(n))} \nonumber \\
&\preceq& \gamma_{max} \frac{(1-\rho(n))(-\log (1-\rho(n))+1)}{\log^2 (1-\rho(n))} \nonumber \\
&\equiv& \frac{\rho(n)(-\log (1-\rho(n))+1)}{\sqrt{n}\log^2 (1-\rho(n))} \preceq 1
\end{eqnarray}

Therefore, the links directly connected to the server will never be a bottleneck. On the other hand, the traffic carried by a node to cache content in level $j$ is $\sum_{i=1}^{\sqrt{n}-j}\gamma i(1-\rho(n))^i \preceq \sum_{i=1}^{\sqrt{n}}\gamma i(1-\rho(n))^i$, so the server links carry the maximum load, and thus the derived capacity is supportable in every link. 

Scenario $\romannumeral 2$- Each one of the four links connected to the server will carry all the traffic related to the items not found in any caches closer to the requester. Thus, the total traffic carried by each of those links is

\begin{eqnarray}
&\gamma (1-\rho(n))+\sum_{i=1}^{\sqrt{n}}4\gamma i (1-\rho(n))^{(1+4\sum_{j=1}^ij)}& \nonumber \\
&\equiv \gamma (1-\rho(n))+\sum_{i=1}^{\sqrt{n}}\gamma i (1-\rho(n))^{2i^2+2i+1},& \nonumber \\
&\equiv \gamma\{(1-\rho(n)) +& \nonumber \\
& \frac{(1-\rho(n))^n-(1-\rho(n))^4 }{\log (1-\rho(n))/(1-\rho(n))}+& \nonumber \\
& \frac{\sqrt{-\frac{\log (1-\rho(n))}{1-\rho(n)}}(erf(\sqrt{-n\log (1-\rho(n))})-erf(\sqrt{-\log (1-\rho(n))}))}{\log (1-\rho(n))/(1-\rho(n))}\}. & \nonumber \\
\end{eqnarray} 

If $\rho(n) \preceq \frac{1}{n}$, then $(1-\rho(n))^{2i^2+2i+1} \equiv 1$ for all $1\leq i \leq \sqrt{n}$. Thus the above traffic will be $n\gamma \preceq n\gamma_{max} \equiv 1$.

If $\rho(n) \succeq \frac{1}{n}$ and $\rho(n) \rightarrow 0$, then using $\log (1-\rho(n)) \equiv -\rho(n)$ the above equation is equivalent to $\frac{\gamma}{\rho(n)} \preceq \frac{1}{\sqrt{n\rho(n)}}$, which is less than $1$ in order.

Finally, if $\rho(n)=\Theta(1)$, then the traffic is equivalent to $\gamma$, which is less than $\Theta(1)$.
 So server links will not be a bottleneck. Using similar reasoning as in scenario $\romannumeral 2$ other links carry less traffic, so the derived capacities are supportable.

Scenario $\romannumeral 3$- The traffic load between the server cell and each of the four neighbor cells is given by

\begin{eqnarray}
&&\gamma \log (n) \{(1-\rho(n))+\sum_{i=2}^{\sqrt{\frac{n}{\log n}}}i(1-\rho(n))^{i\log n}\} \nonumber \\
&\equiv& \gamma \log (n) \{(1-\rho(n))  \nonumber \\
&& +\frac{(1-\rho(n))^{\sqrt{n\log n}}(\sqrt{n\log n}\log (1-\rho(n))-1)}{\log^2 (1-\rho(n))^{\log n}} \nonumber \\
&& -\frac{(1-\rho(n))^{\log n}(\log (1-\rho(n))^{\log n}-1)}{\log^2 (1-\rho(n))^{\log n}}\} \nonumber \\
\label{eq44}
\end{eqnarray}

If $\rho(n) \preceq \frac{1}{\sqrt{n\log n}}$, then $(1-\rho(n))^{i\log n}\rightarrow 1$ for $2\leq i \leq \sqrt{\frac{n}{\log n}}$, thus the traffic load equals to $\gamma \log n \sum_{i=2}^{\sqrt{\frac{n}{\log n}}}i \equiv n\gamma \equiv \sqrt{\frac{n}{\log n}} \succ 1$. Therefore, the obtained capacity is not supported for very small $\rho(n)$ ($\preceq \frac{1}{\sqrt{n\log n}}$). The maximum supportable throughput capacity in this case is $\gamma \preceq \frac{1}{n}$.  

If $\frac{1}{\sqrt{n\log n}} \preceq \rho(n) \preceq \frac{1}{\log n}$, then the maximum traffic load on a link is 

\begin{eqnarray}
\gamma \log n + \gamma \log n \frac{1 + 2 \rho(n) \log n}{\rho^2(n) \log^2 n} 
\equiv \frac{\gamma}{\rho^2(n)\log n}
\end{eqnarray}

The maximum throughput capacity obtained for this region is $\Theta(\rho(n))$, which will lead to a traffic load of $\frac{1}{\rho(n) \log n} \succeq 1$, which means that this bit rate is not supportable. The maximum supportable rate in this region is then $\rho^2(n)\log n$, which is much less than $\rho(n)$.

If $\rho(n) \succeq \frac{1}{\log n}$ and $\rho(n) \rightarrow 0$, then equation \eqref{eq44}  is equivalent to $\gamma \log n \equiv 1$, which is supportable. If $\rho(n) \succeq \frac{1}{\log n}$ and $\rho(n) \nrightarrow 0$, then the maximum traffic is $\gamma$, which is less than $1$, and supportable.

Note that if there were no cache in the system, or $\rho(n)$ is very low, less than the stated threshold values, almost all the requests would be served by the server, and the maximum download rate would be $\Theta(\frac{W}{n})$ in case $\romannumeral 1$, $\romannumeral 2$ and $\Theta(\frac{W}{\sqrt{n\log n}})$ in case $\romannumeral 3$.

\end{IEEEproof}

The maximum throughput capacity is the value which can be supported by all the nodes while no interference is occurred. Thus combining Lemmas \ref{lem:03} and \ref{lem:04}, Theorem \ref{thm:03} is proved.

\section{Discussion}
\label{sec:discussion}

In this section, we discuss our results based on two examples. The first example is that of a grid wireless network with $n$ caches, and one server, which contains all the items located in the middle of the network. The requesters use the path search to locate the contents.  In the second example we study the impact of caching on the maximum capacity order in the grid and random networks where all the caches have the same probability of having each item at any given time. The networks where the received data is stored only at the receivers and then shared with the other nodes as long as the node keeps the content can be considered as an example of such networks.

\subsubsection{Example 1}
\label{ex:01}

Assume that each cache in level $i$ (nodes at $i$ hops away from the server) in a grid network receives requests for a specific document according to a Poisson distribution with rate $\beta$ from the local user, and with rate $\beta'_i(n)$ from all the other nodes. Note that rate $\beta'_i(n)$ is a function of the individual request rate of users ($\beta$) and also the location of the cache in the network. The content discovery mechanism is path-wise discovery, and whenever a copy of the required data is found (in a cache or server), it will be downloaded through the reverse path, and all the nodes on the download path store it in their local caches. Moreover, we assume that receiving the data and also any request for the available cached data by a node in level $i$ refreshes a time-out timer with fixed duration $D_i$. According to \cite{Che2002Hierarchical},  this is a good approximation for caches with Least Recently Used (LRU) replacement policy when the cache size and the total number of documents are reasonably large. We will calculate the average probability of the data being in a cache in level $i$ ($\rho_i(n)$) based on these assumptions and then use Theorem \ref{thm:01} to obtain the throughput capacity.

Let random variable $t_{on}(T)$ denote the total time of the data being available in a cache during constant time $T$. Assume that $N(T)$ requests are received by each node $v_i$ in level $i$ ($i$ hop distance from the server). The data available time between any two successive requests (internal and external) is $D_i$ if the timer set by the first request is expired before the second one comes, or is equal to the time between these two requests. Let $\tau^{req}_i$ denote the time between receiving two successive requests. This process has an exponential distribution with parameter $\beta_i=\beta+\beta'_i$. So the total time of data availability in a level $i$ cache is 

\begin{eqnarray}
t_{on}(T)=\sum_{k=0}^{N(T)} \min(\tau_i^{req},D_i),
\end{eqnarray} 
and the average value of this time is 
\begin{eqnarray}
E[t_{on}(T)]&=&\sum_{m=0}^{\infty} E[\sum_{k=0}^{m} \min(\tau_i^{req},D_i)]Pr(N(T)=m), \nonumber \\
&=&\sum_{m=0}^{\infty} mE[\min(\tau_i^{req},D_i)]Pr(N(T)=m), \nonumber \\
&=&E[\min(\tau_i^{req},D_i)]E[N(T)].
\end{eqnarray}

According to the Poisson arrivals of requests with parameter $\beta+\beta'_i$, $E[N(T)]=(\beta+\beta'_i)T$. 

$E[\min(\tau_i^{req},D_i)]$ can be easily calculated and equals to $\frac{1-e^{-D_i(\beta+\beta'_i)}}{\beta+\beta'_i}$. Therefore,

\begin{eqnarray}
E[t_{on}(T)]&=&(1-e^{-D_i(\beta+\beta'_i)})T
\end{eqnarray}
 
And finally the probability of an item being available in a level $i$ cache is $\rho_i=\frac{E[t_{on}(T)]}{T}=1-e^{-D_i(\beta+\beta'_i(n))}$. 
Note that $D_0=\infty$ so that $\rho_0=1$.

Now we need to calculate the rate of requests received by each node in level $i$. We assume that the shortest path from the requester to the server is selected such that all the nodes in level $i$ receive the requests with the same rate. There are $4i$ nodes in level $i$ and $4(i+1)$ nodes in level $i+1$. So the request initiated or forwarded from a node in level $i+1$ will be received by a specific node in level $i$ with probability $\frac{i}{i+1}$ if it is not locally available in that node, so $\beta'_i(n)$ can be expressed as

\begin{eqnarray}
\beta'_i=\frac{(1-\rho_{i+1})(\beta+\beta'_{i+1})(i+1)}{i}  \label{eq:beta}
\end{eqnarray} 

Combining equation \ref{eq:beta}, the relationship between $\rho_i$ and $\beta'_i$, and the fact that there is no external request coming to the nodes at the edge boundary of the network ($\beta'_{\sqrt{n}}=\beta$), together with the result of Theorem \ref{thm:01} we can obtain the capacity ($\gamma_{max}$) in the grid network with path-wise content discovery and on-path storing scheme which is given by

\begin{eqnarray}
\frac{W\sqrt{n}/4}{\sum_{i=1}^{\sqrt{n}}i\sum_{j=0}^i e^{-\sum_{k=j+1}^i D_k(\beta+\beta'_k)}(1-e^{-D_j(\beta+\beta'_j)})} \nonumber \\
\end{eqnarray}  

\begin{figure}[http]
    \center
       \includegraphics[scale=0.23,angle=0]{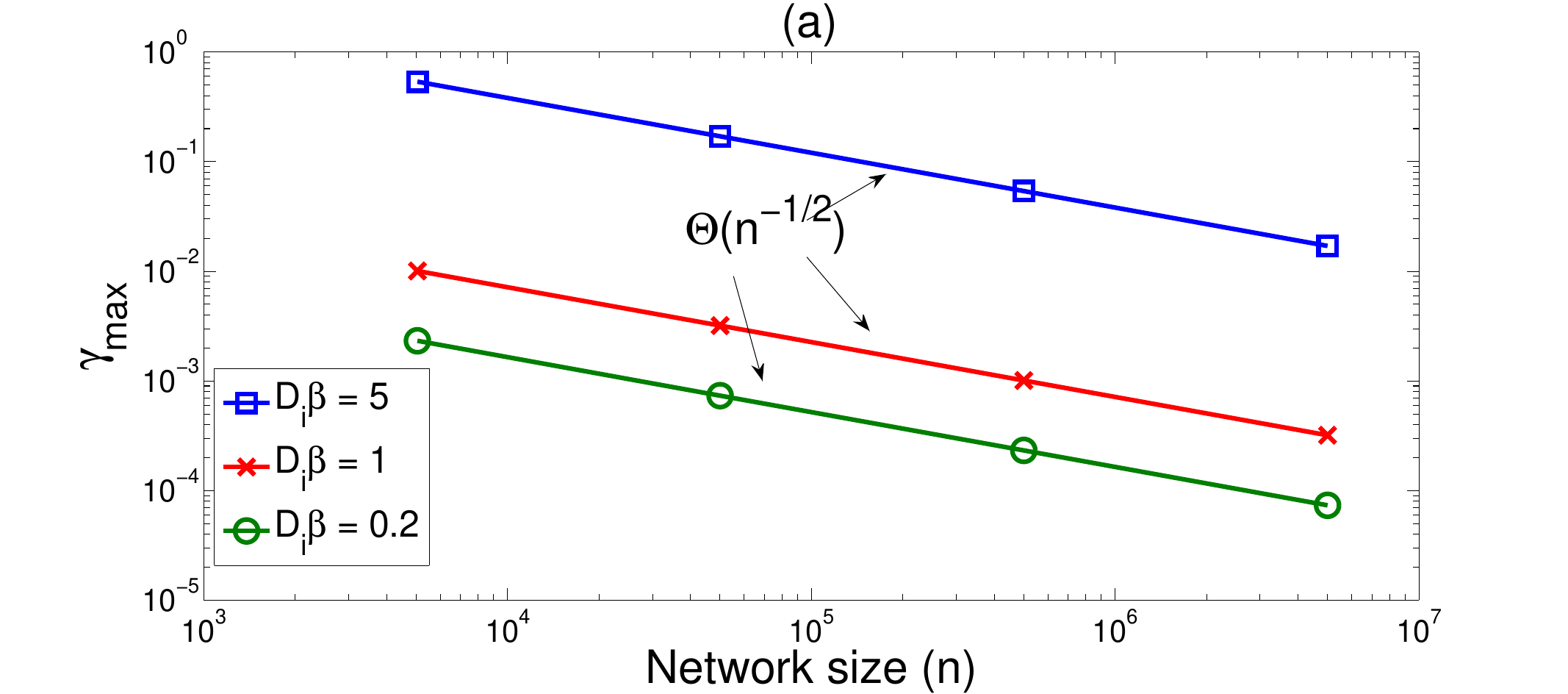}\\
			 \includegraphics[scale=0.23,angle=0]{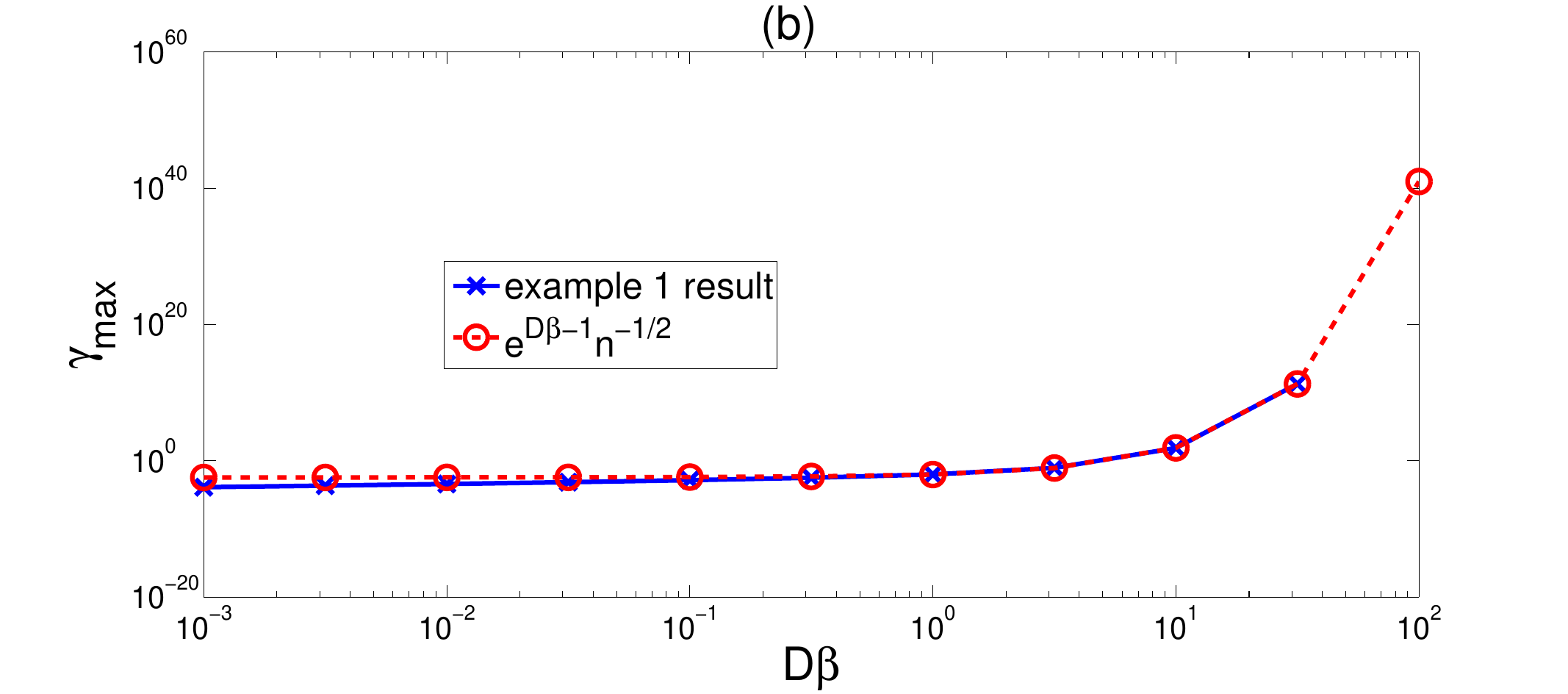}\\
      \caption{Maximum throughput capacity ($\gamma_{max}$) versus (a) network size ($n$), (b) Timeout-request rate product ($\beta D$).}
    \label{fig:onpath}
\end{figure}

Figure \ref{fig:onpath} (a) illustrates that the maximum throughput capacity changes with the network size ($n$) when $D_i\beta$ is the same for all nodes. It can be seen that the this capacity is inversely proportional to $\sqrt{n}$, just like the throughput capacity when no timer refreshing is available and the downloaded data is stored just in the end user's cache. 

Figure \ref{fig:onpath} (b) shows the capacity versus different values for $D_i\beta$ assuming $n=10^4$ and same timer expiration time for all nodes. It can be seen that the maximum capacity is very close to $e^{D\beta-1}/\sqrt{n}$. For large $D\beta$ products the probability of the content being available in each and every cache will tend to be one, so all the contents are downloaded from the local cache and no data transfer is needed to be done, therefore the calculated throughput capacity will be very large which means that the all the links are available with their maximum bandwidth.

\subsubsection{Example 2}
\label{ex:02}

As a possible example leading to equal probability of all the caches containing a piece of data, which is the basic assumption of Theorems \ref{thm:02} and \ref{thm:03}, assume that receiving a data in the local cache of the requesting user sets a time-out timer with exponentially distributed duration with parameter $\eta$ and no other event will change the timer until it times-out, meaning that $\mu=\eta$. Considering the request process for each content from each user being a Poisson process with rate $\beta$, and using the memoryless property of exponential distribution (internal request inter-arrival times), and assuming that the received data is stored only in the end user's cache (the caches on the download path don't store the downloading data), it can be proved that $\lambda=\beta$. Thus we can write the presence probability of each content in each cache as $\rho(n)=\frac{\beta}{\beta+\eta}$. 

\begin{figure}[http]
    \center
			\includegraphics[scale=0.23,angle=0]{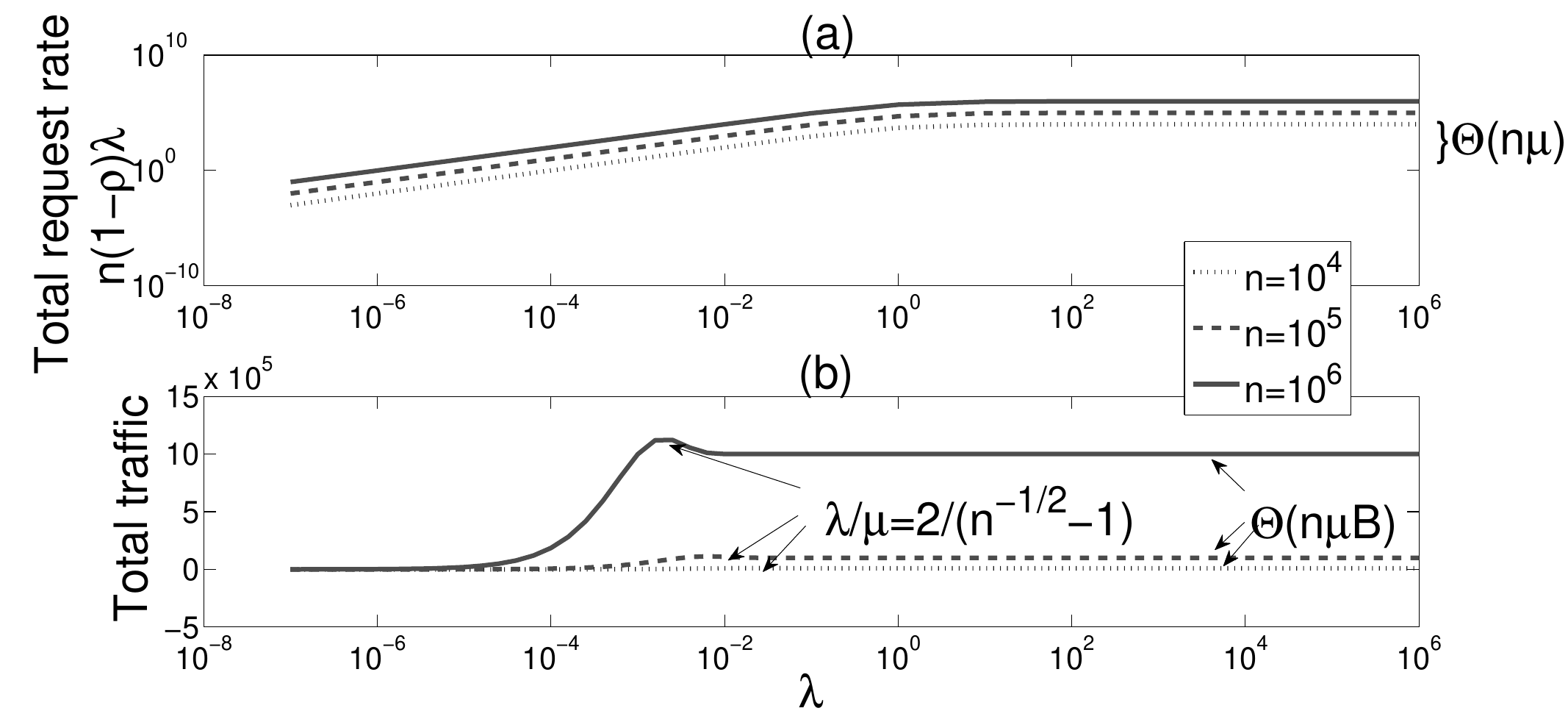}\\
      \caption{(a) Total request rate in the network ($\lambda n (1-\rho(n))$), (b) Total traffic in the network ($B\lambda n (1-\rho(n))E[h]$) vs. the request rate ($\lambda$) with fixed time-out rate ($\mu=1$).}
    \label{fig:traffic_lambda}
\end{figure}

Figures \ref{fig:traffic_lambda} (a),(b) respectively illustrate the total request rate and the total traffic generated in a fixed size network in scenario $\romannumeral 1$ for different request rates when the time-out rate is fixed. The total request rate in the network is the product of the number of requesting nodes and the rate at which each node is sending the request ($n(1-\rho)\lambda$). The total traffic is the product of the total request rate and the number of hops between source and destination and the content size ($n(1-\rho)\lambda B E[h]$).
Small $\lambda$ means that each node is sending requests with low rate, so fewer caches have the content, and consequently more nodes are sending requests with this low rate. In this case most of the requests are served by the server. The total request rate will increase by increasing the per node request rate. High $\lambda$ shows that each node is requesting the content with higher rate, so the number of cached content in the network is high, thus fewer nodes are requesting the content with this high rate externally. Here most of the requests are served by the caches. The total request rate then is determined by the content drop rate. So for very large $\lambda$, the total request rate is the total number of nodes in the network times the drop rate ($n\mu$) and the total traffic is $n\mu B$. As can be seen there is some request rate at which the traffic reaches its maximum; this happens when there is a balance between the requests served by the server and by the caches, for smaller request rates, most of the requests are served by the server and increasing $\lambda$ increases the total traffic; for larger $\lambda$, on the other hand, most of the requests are served by the caches and increasing the request rate will decrease the distance to the nearest content and decrease total traffic.

\begin{figure}[http]
    \center
			\includegraphics[scale=0.23,angle=0]{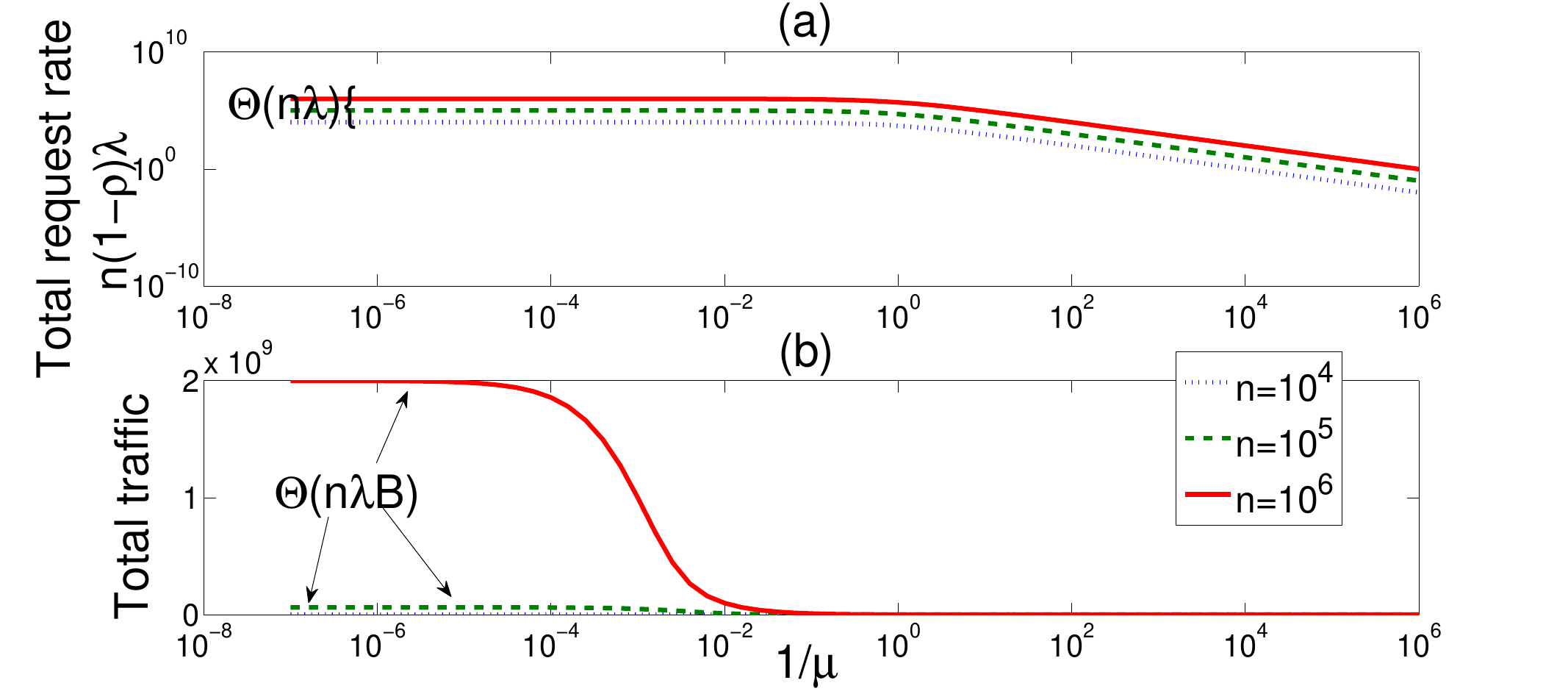}\\
      \caption{(a) Total request rate in the network ($\lambda n (1-\rho(n))$), (b) Total traffic in the network ($B\lambda n (1-\rho(n))E[h]$) vs. the inverse of the time-out rate ($1/\mu$) with fixed request ratio ($\lambda=1$).}
    \label{fig:traffic_mu}
\end{figure}

Figures \ref{fig:traffic_mu} (a),(b) respectively illustrate the total request rate and the total traffic generated in a fixed size network in scenario $\romannumeral 1$ for different time-out rates when the request rate is fixed. Low $1/\mu$ means high time-out rates or small lifetimes, which means most of the requests are served by the server and caching is not used at all. For large time-out times, all the requests are served by the caches, and the only parameter in determining the total request rate is the time-out rate.

However, when the network grows the traffic in the network will increase and the download rate will decrease. If we assume that the new requests are not issued in the middle of the previous download, the request rate will decrease with network growth. If the holding time of the contents in a cache increases accordingly the total traffic will not change, i.e. if by increasing the network size the requests are issued not as fast as before, and the contents are kept in the caches for longer times, the network will perform similarly.

\begin{figure}[http]
    \center
			\includegraphics[scale=0.23,angle=0]{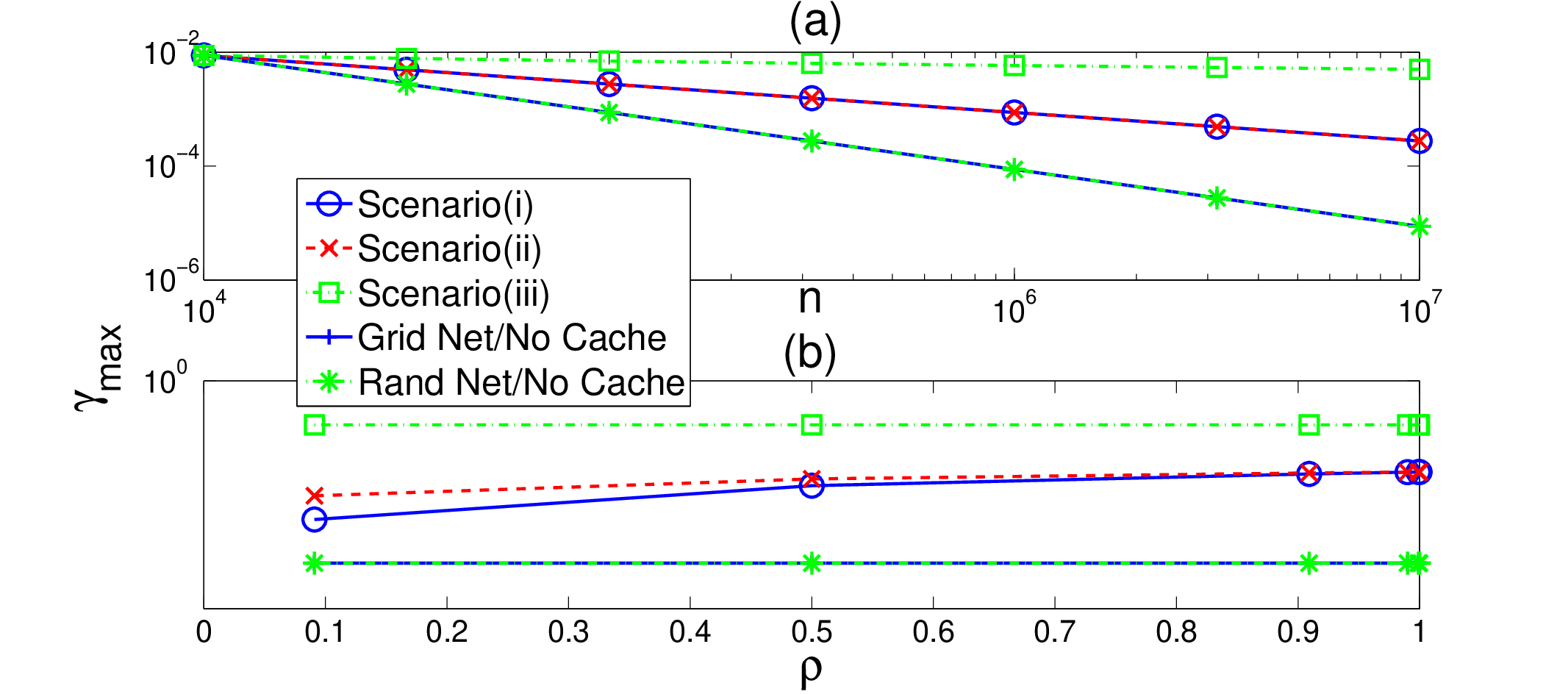}\\
      \caption{Maximum download rate ($\gamma_{max}$) vs. (a) the number of nodes ($n$), (b) the content presence probability($\rho(n)$).}
    \label{fig:capacity}
\end{figure}

In Figure \ref{fig:capacity} (a) we assume that the request rate is roughly $7$ times the drop rate, so $\rho(n)=7/8$, and show the maximum throughput order as a function of the network size. According to Theorem \ref{thm:02} and as can be observed from this figure, the maximum throughput capacity of the network in a grid network with the described characteristics is inversely proportional to the square root of the network size if the probability of each item being in each cache is fixed, while in a network with no cache this capacity will be inversely proportional to the network size. Similarly in the random network the maximum throughput is inversely proportional to the logarithm of the network size.

Moreover, comparing scenario $\romannumeral 1$ with $\romannumeral 2$, we observe that the throughput capacity in both cases are almost the same; meaning that using the path discovery scheme will lead to almost the same throughput capacity as the expanding ring discovery. Thus, we can conclude that just by knowing the address of a server containing the required data and forwarding the requests through the shortest path toward that server we can achieve the best performance, and increasing the complexity and control traffic to discover the closest copy of the required content does not add much to the capacity. 

On the other hand with a fixed network size, if the probability of an item being in each cache is greater than a threshold ($\Theta(\frac{1}{\sqrt{n}})$, $\Theta(\frac{1}{n})$, and $\Theta(\frac{1}{\log n})$ in cases $\romannumeral 1,\romannumeral 2$ and $\romannumeral 3$, respectively), most of the requests will be served by the caches and not the server, so increasing the probability of an intermediate cache having the content reduces the number of hops needed to forward the content to the customer, and consequently increases the throughput (Figure \ref{fig:capacity} (b), $n=10^4$). For content presence probability orders less than these thresholds ($\Theta(\frac{1}{\sqrt{n\log n}})$ in cases $\romannumeral 3$) most of the requests are served by the main server, so the maximum possible number of hops will be traveled by each content to reach the requester and the minimum throughput capacity ($\Theta(\frac{W}{n})$) will be achieved. Note that in random network, the maximum throughput is limited by the maximum supportable load on each link.

As may have been expected and according to our results, the obtained throughput is a function of the probability of each content being available in each cache, which in turn is strongly dependent on the network configuration and cache management policy. 

\section{Conclusion And Future Work}
\label{sec:conclusion}

We studied the asymptotic throughput capacity and latency of ICNs with limited lifetime cached data at each node. The grid and random networks are two network models we investigated in this work. The results show that with fixed content presence probability in each cache, the network can have the maximum throughput order of $1/\sqrt{n}$ and $1/\log n$ in cases of grid and random networks, respectively, and the number of hops travelled by each data to reach the customer (or latency of obtaining data), can be as small as one hop. 


Moreover, we studied the impact of the content discovery mechanism on the performance. It can be observed that looking for the closest cache containing the content will not have much asymptotic advantage over the simple path-wise discovery. Consequently, downloading the nearest available copy on the path toward the server will have the same performance as downloading from the nearest copy. A practical consequence of this result is that routing may not need to be updated with knowledge of local copies, just getting to the source and finding the content opportunistically will yield the same benefit. 

Another interesting finding is that whether all the caches on the download path keep the data or just the end user does it, the maximum throughput capacity scale does not change. 

In this work, we have made several assumptions to simplify the analysis. For example, we assumed all the contents have the same characteristics (size, popularity). This assumption should be relaxed in future work. We also assumed that the requester downloads the data completely from one content location. However, if the node that needs the data can download each part of it from different nodes and makes a complete content out of the collected parts, achievable capacities may be different. Proposing a caching and downloading scheme that can improve the capacity order is part of our  future work.

\bibliographystyle{IEEEtran}
\bibliography{IEEEabrv,ToN2015}

\begin{thebibliography}{10}
\providecommand{\url}[1]{#1}
\csname url@samestyle\endcsname
\providecommand{\newblock}{\relax}
\providecommand{\bibinfo}[2]{#2}
\providecommand{\BIBentrySTDinterwordspacing}{\spaceskip=0pt\relax}
\providecommand{\BIBentryALTinterwordstretchfactor}{4}
\providecommand{\BIBentryALTinterwordspacing}{\spaceskip=\fontdimen2\font plus
\BIBentryALTinterwordstretchfactor\fontdimen3\font minus
  \fontdimen4\font\relax}
\providecommand{\BIBforeignlanguage}[2]{{%
\expandafter\ifx\csname l@#1\endcsname\relax
\typeout{** WARNING: IEEEtran.bst: No hyphenation pattern has been}%
\typeout{** loaded for the language `#1'. Using the pattern for}%
\typeout{** the default language instead.}%
\else
\language=\csname l@#1\endcsname
\fi
#2}}
\providecommand{\BIBdecl}{\relax}
\BIBdecl

\bibitem{Zhang2010Named}
L.~Zhang, D.~Estrin, J.~Bruke, V.~Jacobson, J.~Thornton, D.~Smetters, B.~Zhang,
  G.~Tsudik, K.~Claffy, D.~Krioukov, D.~Massey, C.~Papadopoulos, T.~Abdelzaher,
  L.~Wang, P.~Crowley, and E.~Yeh, ``Named data networking ({NDN}) project,''
  Oct. 2010.

\bibitem{Pursuit}
``{PURSUIT}: Pursuing a pub/sub internet,'' http://www.fp7-pursuit.eu/, Sep.
  2010.

\bibitem{Ahlgren2012Survey}
B.~Ahlgren, C.~Dannewitz, C.~Imbrenda, D.~Kutscher, and B.~Ohlman, ``A survey
  of information-centric networking,'' \emph{Communications Magazine, IEEE},
  vol.~50, no.~7, July 2012.

\bibitem{Jacobson2009Networking}
V.~Jacobson, D.~K. Smetters, J.~D. Thornton, M.~F. Plass, N.~H. Briggs, and
  R.~L. Braynard, ``Networking named content,'' in \emph{ACM CoNEXT}, 2009, pp.
  1--12.

\bibitem{Azimdoost2013Throughput}
B.~Azimdoost, C.~Westphal, and H.~R. Sadjadpour, ``{On the throughput capacity
  of information-centric networks},'' in \emph{Teletraffic Congress (ITC), 2013
  25th International}.\hskip 1em plus 0.5em minus 0.4em\relax IEEE, 2013, pp.
  1--9.

\bibitem{Ahlgren2008Design}
B.~Ahlgren, M.~D'Ambrosio, M.~Marchisio, I.~Marsh, C.~Dannewitz, B.~Ohlman,
  K.~Pentikousis, O.~Strandberg, R.~Rembarz, and V.~Vercellone, ``Design
  considerations for a network of information,'' in \emph{ACM CoNEXT}, 2008,
  pp. 1--6.

\bibitem{Koponen2007Dataoriented}
T.~Koponen, M.~Chawla, B.~G. Chun, A.~Ermolinskiy, K.~H. Kim, S.~Shenker, and
  I.~Stoica, ``A data-oriented (and beyond) network architecture,'' in
  \emph{ACM SIGCOMM}, 2007, pp. 181--192.

\bibitem{Ghodsi2011InformationCentric}
A.~Ghodsi, T.~Koponen, B.~Raghavan, S.~Shenker, A.~Singla, and J.~Wilcox,
  ``{Information-Centric} networking: Seeing the forest for the trees,'' in
  \emph{HotNets}, 2011.

\bibitem{Olmos2014Catalog}
F.~Olmos, B.~Kauffmann, A.~Simonian, and Y.~Carlinet, ``{Catalog Dynamics:
  Impact of Content Publishing and Perishing on the Performance of a LRU
  Cache},'' Sep. 2014.

\bibitem{InfoCom01:Che}
H.~Che, Z.~Wang, and Y.~Tung, ``Analysis and design of hierarchical web caching
  systems,'' in \emph{IEEE INFOCOM}, 2001, pp. 1416--1424.

\bibitem{InfoCom10:Rosensweig}
E.~Rosensweig, J.~Kurose, and D.~Towsley, ``Approximate models for general
  cache networks,'' in \emph{IEEE INFOCOM}, 2010, pp. 1--9.

\bibitem{Wolman1999Scale}
A.~Wolman, M.~Voelker, N.~Sharma, N.~Cardwell, A.~Karlin, and H.~M. Levy, ``{On
  the scale and performance of cooperative Web proxy caching},'' \emph{SIGOPS
  Oper. Syst. Rev.}, vol.~33, no.~5, pp. 16--31, Dec. 1999.

\bibitem{Rosensweig2009Breadcrumbs}
E.~J. Rosensweig and J.~Kurose, ``Breadcrumbs: Efficient, {Best-Effort} content
  location in cag networks,'' in \emph{IEEE INFOCOM}, 2009, pp. 2631--2635.

\bibitem{IEEEMob05:Yin}
L.~Yin and G.~Cao, ``Supporting cooperative caching in ad hoc networks,''
  \emph{IEEE Transactions on Mobile Computing}, no.~1, pp. 77--89, 2005.

\bibitem{Tortelli2011Fairness}
M.~Tortelli, I.~Cianci, L.~A. Grieco, G.~Boggia, and P.~Camarda, ``A fairness
  analysis of content centric networks,'' Nov. 2011.

\bibitem{Westphal2005Maximizing}
C.~Westphal, ``On maximizing the lifetime of distributed information in ad-hoc
  networks with individual constraints,'' in \emph{ACM MobiHoc}, 2005, pp.
  26--33.

\bibitem{Dehghan2014Complexity}
M.~Dehghan, A.~Seetharam, B.~Jiang, T.~He, T.~Salonidis, J.~Kurose, D.~Towsley,
  and R.~Sitaraman, ``{On the Complexity of Optimal Routing and Content Caching
  in Heterogeneous Networks},'' Dec. 2014.

\bibitem{Gupta2000Capacity}
P.~Gupta and P.~Kumar, ``The capacity of wireless networks,'' \emph{IEEE
  Transactions on Information Theory}, vol.~46, no.~2, 2000.

\bibitem{Li2001Capacity}
J.~Li, C.~Blake, D.~S. De~Couto, H.~I. Lee, and R.~Morris, ``Capacity of ad hoc
  wireless networks,'' in \emph{MobiCom}, 2001, pp. 61--69.

\bibitem{Niesen2009Capacity}
U.~Niesen, P.~Gupta, and D.~Shah, ``On capacity scaling in arbitrary wireless
  networks,'' \emph{Information Theory, IEEE Transactions on}, vol.~55, no.~9,
  pp. 3959--3982, 2009.

\bibitem{Grossglauser2002Mobility}
M.~Grossglauser and D.~Tse, ``Mobility increases the capacity of ad hoc
  wireless networks,'' \emph{Networking, IEEE/ACM Transactions On}, vol.~10,
  no.~4, pp. 477--486, 2002.

\bibitem{Herdtner2005Throughput}
J.~D. Herdtner and E.~K. Chong, ``Throughput-storage tradeoff in ad hoc
  networks,'' in \emph{IEEE INFOCOM}, 2005, pp. 2536--2542.

\bibitem{AlfanoContentCentric}
G.~Alfano, M.~Garetto, and E.~Leonardi, ``{Content-Centric Wireless Networks
  With Limited Buffers: When Mobility Hurts},'' \emph{IEEE/ACM Transactions on
  Networking}, vol.~20, Oct. 2014.

\bibitem{ICCW09:Liu}
H.~Liu, Y.~Zhang, and D.~Raychaudhuri, ``Performance evaluation of the
  cache-and-forward ({CNF}) network for mobile content delivery services,'' in
  \emph{{ICC} Workshop}, 2009, pp. 1--5.

\bibitem{Carofiglio2011Modeling}
G.~Carofiglio, M.~Gallo, L.~Muscariello, and D.~Perino, ``{Modeling data
  transfer in content-centric networking},'' in \emph{IEEE Teletraffic Congress
  (ITC23)}, 2011, pp. 111--118.

\bibitem{IEEEIT11:Niesen}
U.~Niesen, D.~Shah, and G.~Wornell, ``Caching in wireless networks,''
  \emph{IEEE Transactions on Information Theory}, 2011.

\bibitem{Gitzenis2013Asymptotic}
S.~Gitzenis, G.~S. Paschos, and L.~Tassiulas, ``{Asymptotic Laws for Joint
  Content Replication and Delivery in Wireless Networks},'' \emph{Information
  Theory, IEEE Transactions on}, vol.~59, no.~5, pp. 2760--2776, May 2013.

\bibitem{book06:Xue}
F.~Xue and P.~Kumar, \emph{Scaling Laws for Ad Hoc Wireless Networks: an
  Information Theoretic Approach}.\hskip 1em plus 0.5em minus 0.4em\relax
  Foundations and Trends in Networking, {NOW} Publishers, 2006.

\bibitem{Applied97:Penrose}
M.~D. Penrose, ``The longest edge of the random minimal spanning tree,''
  \emph{The Annals of Applied Probability}, pp. 340--361, 1997.

\bibitem{Che2002Hierarchical}
H.~Che, Y.~Tung, and Z.~Wang, ``{Hierarchical Web caching systems: modeling,
  design and experimental results},'' \emph{IEEE Journal on Selected Areas in
  Communications}, vol.~20, no.~7, pp. 1305--1314, Sep. 2002.

\end{thebibliography}

\end{document}